\documentclass{article}%
\usepackage{amsmath}
\usepackage{amsfonts}
\usepackage{amssymb}%
\usepackage{graphicx}
\begin{document}

\title{The Cosmological Constant in Distorted Quantum Cosmology}
\author{Remo Garattini\thanks{Frontiers of Fundamental Physics 14, 15-18 July
2014,\newline\indent  Aix Marseille University (AMU) Saint-Charles Campus, Marseille,
France}\\University of Bergamo, \\Department of Engineering and Applied Sciences\\Viale Marconi 5, 24044 Dalmine (Bergamo) Italy\\I.N.F.N. - sezione di Milano, Milan, Italy\\E-mail: remo.garattini@unibg.it}
\date{}
\maketitle

\begin{abstract}
We give a calculation scheme for the cosmological constant computation with
the help of the Wheeler-DeWitt equation. This last one is regarded as a
Sturm-Liouville problem with the cosmological constant considered as the
associated eigenvalue. By fixing the ideas on a
Friedmann-Lema\^{\i}tre-Robertson-Walker line element in ordinary gravity, we
apply this calculation scheme on distorted gravity. By distorted gravity, we
mean all the deviations from General Relativity. We restrict our proposal on
Gravity's Rainbow and Noncommutative geometry. A brief comment on
Ho\v{r}ava-Lifshitz (HL) theory is discussed.

\end{abstract}

\section{Introduction}

\label{p1}

One of the cornerstone of Quantum Cosmology is the Wheeler-DeWitt (WDW)
equation\cite{DeWitt}. It represents the quantum version of the invariance
with respect to time reparametrization. If we denote with $d\Omega_{3}%
^{2}=\gamma_{ij}dx^{i}dx^{j}$ the line element on the three-sphere, with $N$
the lapse function and with $a(t)$ the scale factor, the
Friedmann-Lema\^{\i}tre-Robertson-Walker line element assumes the form%
\begin{equation}
ds^{2}=-N^{2}dt^{2}+a^{2}\left(  t\right)  d\Omega_{3}^{2}.\label{FRW}%
\end{equation}
Thus the WDW in absence of matter fields is%
\begin{align}
H\Psi\left(  a\right)   & =\left[  -a^{-q}\left(  \frac{\partial}{\partial
a}a^{q}\frac{\partial}{\partial a}\right)  +\frac{9\pi^{2}}{4G^{2}}\left(
a^{2}-\frac{\Lambda}{3}a^{4}\right)  \right]  \Psi\left(  a\right) \nonumber\\
& =\left[  -\frac{\partial^{2}}{\partial a^{2}}-\frac{q}{a}\frac{\partial
}{\partial a}+\frac{9\pi^{2}}{4G^{2}}\left(  a^{2}-\frac{\Lambda}{3}%
a^{4}\right)  \right]  \Psi\left(  a\right)  =0,\label{WDW_0}%
\end{align}
where we have also introduced a factor order ambiguity $q$, the Newton's
constant $G$ and a cosmological constant $\Lambda$. In this formulation all
the degrees of freedom except the scale factor $a\left(  t\right)  $ have been
integrated. If the WDW equation is interpreted as an eigenvalue equation, one
simply finds%
\begin{equation}
H\Psi\left(  a\right)  =E\Psi\left(  a\right)  =0,
\end{equation}
namely a zero energy eigenvalue. However, it appears that the WDW equation has
also a hidden structure. Indeed Eq.(\ref{WDW_0}) has the structure of a
Sturm-Liouville eigenvalue problem with the cosmological constant as
eigenvalue. We recall to the reader that a Sturm-Liouville differential
equation is defined by%
\begin{equation}
\frac{d}{dx}\left(  p\left(  x\right)  \frac{dy\left(  x\right)  }{dx}\right)
+q\left(  x\right)  y\left(  x\right)  +\lambda w\left(  x\right)  y\left(
x\right)  =0\label{SL}%
\end{equation}
and the normalization is defined by%
\begin{equation}
\int_{a}^{b}dxw\left(  x\right)  y^{\ast}\left(  x\right)  y\left(  x\right)
,
\end{equation}
where the boundary conditions are momentarily suspended. It is a standard
procedure, to convert the Sturm-Liouville problem (\ref{SL}) into a
variational problem of the form%
\begin{equation}
F\left[  y\left(  x\right)  \right]  =\frac{-\int_{a}^{b}dxy^{\ast}\left(
x\right)  \left[  \frac{d}{dx}\left(  p\left(  x\right)  \frac{d}{dx}\right)
+q\left(  x\right)  \right]  y\left(  x\right)  }{\int_{a}^{b}dxw\left(
x\right)  y^{\ast}\left(  x\right)  y\left(  x\right)  }\,,\label{Funct}%
\end{equation}
with unspecified boundary condition. If $y\left(  x\right)  $ is an
eigenfunction of (\ref{SL}), then%
\begin{equation}
\lambda=\frac{-\int_{a}^{b}dxy^{\ast}\left(  x\right)  \left[  \frac{d}%
{dx}\left(  p\left(  x\right)  \frac{d}{dx}\right)  +q\left(  x\right)
\right]  y\left(  x\right)  }{\int_{a}^{b}dxw\left(  x\right)  y^{\ast}\left(
x\right)  y\left(  x\right)  }\,,
\end{equation}
is the eigenvalue, otherwise%
\begin{equation}
\lambda_{1}=\min_{y\left(  x\right)  }\frac{-\int_{a}^{b}dxy^{\ast}\left(
x\right)  \left[  \frac{d}{dx}\left(  p\left(  x\right)  \frac{d}{dx}\right)
+q\left(  x\right)  \right]  y\left(  x\right)  }{\int_{a}^{b}dxw\left(
x\right)  y^{\ast}\left(  x\right)  y\left(  x\right)  }\,.
\end{equation}
\textbf{\ }The minimum of the functional $F\left[  y\left(  x\right)  \right]
$ corresponds to a solution of the Sturm-Liouville problem (\ref{SL}) with the
eigenvalue $\lambda.$ In the case of the FLRW model we have the following
correspondence%
\begin{align}
p\left(  x\right)   & \rightarrow a^{q}\left(  t\right)  \,,\nonumber\\
q\left(  x\right)   & \rightarrow\left(  \frac{3\pi}{2G}\right)  ^{2}%
a^{q+2}\left(  t\right)  \,,\nonumber\\
w\left(  x\right)   & \rightarrow a^{q+4}\left(  t\right)  \,,\nonumber\\
y\left(  x\right)   & \rightarrow\Psi\left(  a\right)  \,,\nonumber\\
\lambda & \rightarrow\frac{\Lambda}{3}\left(  \frac{3\pi}{2G}\right)  ^{2}\,.
\end{align}
Since $a\left(  t\right)  \in\left[  0,\infty\right)  $, the normalization
becomes%
\begin{equation}
\int_{0}^{\infty}daa^{q+4}\Psi^{\ast}\left(  a\right)  \Psi\left(  a\right)
,\label{Norm1}%
\end{equation}
where it is understood that $\Psi\left(  \infty\right)  =0$. In the
minisuperspace approach with a FLRW background, one finds%
\begin{equation}
\frac{\int\mathcal{D}aa^{q}\Psi^{\ast}\left(  a\right)  \left[  -\frac
{\partial^{2}}{\partial a^{2}}-\frac{q}{a}\frac{\partial}{\partial a}%
+\frac{9\pi^{2}}{4G^{2}}a^{2}\right]  \Psi\left(  a\right)  }{\int
\mathcal{D}aa^{q+4}\Psi^{\ast}\left(  a\right)  \Psi\left(  a\right)  }%
=\frac{3\Lambda\pi^{2}}{4G^{2}}.\label{WDW_1}%
\end{equation}
As a concrete case, fixing $q=0$ and taking as a trial wave function
$\Psi\left(  a\right)  =\exp\left(  -\beta a^{2}\right)  $, one finds
$\Psi\left(  a\right)  \rightarrow0$ when $a\rightarrow\infty$. Then the only
solution allowed is complex and therefore it must be discarded\cite{RemoHL}.
Of course, the general $q$ case is much more complicated\cite{RGMdL}. Note
that the global energy eigenvalue is still vanishing. What we can compute in
the Sturm-Liouville formulation is the degree of degeneracy which is
represented by the cosmological constant and the value of the cosmological
constant itself. In the next section we give the general guidelines in
ordinary gravity and in presence of Modified Dispersion Relations and the Non
Commutative approach to Quantum Field Theory. Units in which $\hbar=c=k=1$ are
used throughout the paper.

\section{The Cosmological Constant in Distorted Quantum Cosmology}

When we generalize the minisuperspace approach of Eq.(\ref{WDW_1}), the formal
structure persists. Indeed the WDW equation can be transformed into
($\kappa=8\pi G$)%
\begin{equation}
\frac{1}{V}\frac{\int\mathcal{D}\left[  g_{ij}\right]  \Psi^{\ast}\left[
g_{ij}\right]  \int_{\Sigma}d^{3}x\hat{\Lambda}_{\Sigma}\Psi\left[
g_{ij}\right]  }{\int\mathcal{D}\left[  g_{ij}\right]  \Psi^{\ast}\left[
g_{ij}\right]  \Psi\left[  g_{ij}\right]  }=\frac{1}{V}\frac{\left\langle
\Psi\left\vert \int_{\Sigma}d^{3}x\hat{\Lambda}_{\Sigma}\right\vert
\Psi\right\rangle }{\left\langle \Psi|\Psi\right\rangle }=-\frac{\Lambda
}{\kappa},\label{VEVO}%
\end{equation}
where we have integrated over the hypersurface $\Sigma$ and we have defined%
\begin{equation}
V=\int_{\Sigma}d^{3}x\sqrt{g}\label{Vol}%
\end{equation}
as the volume of the hypersurface $\Sigma$ with%
\begin{equation}
\hat{\Lambda}_{\Sigma}=\left(  2\kappa\right)  G_{ijkl}\pi^{ij}\pi^{kl}%
-\sqrt{g}R/\left(  2\kappa\right)  .\label{LambdaSigma}%
\end{equation}
$G_{ijkl}$ is the supermetric, while $\pi^{ij}$ is termed the supermomentum,
$R$ is the three scalar curvature and$\sqrt{g}$ is the determinant of the
three metric $g_{ij}$. In this form, Eq.(\ref{VEVO}) can be used to compute
Zero Point Energy (ZPE) provided that $\Lambda/\kappa$ be considered as an
eigenvalue of $\hat{\Lambda}_{\Sigma}$, namely the WDW equation is transformed
into an expectation value computation. Nevertheless, solving Eq.(\ref{VEVO})
is a quite impossible task, therefore we are oriented to use a variational
approach with trial wave functionals. The related boundary conditions are
dictated by the choice of the trial wave functionals which, in our
case\textbf{,} are of the Gaussian type: this choice is justified by the fact
that ZPE should be described by a good candidate of the \textquotedblleft%
\textit{vacuum state}\textquotedblright. However if we change the form of the
wave functionals we change also the corresponding boundary conditions and
therefore the description of the vacuum state. It is better to observe that
the obtained eigenvalue $\Lambda/\kappa$, it is far to be a constant, rather
it will be dependent on some parameters and therefore it will be considered
more like a \textquotedblleft\textit{dynamical cosmological constant}%
\textquotedblright. Usually, when we compute Eq.(\ref{VEVO}) to one loop or
higher loops, UltraViolet divergences appear. In ordinary gravity, to take
under control such divergencies we need a regularization/renormalization
scheme\cite{Remo}. However, as shown by Ho\v{r}ava, a modification of Einstein
gravity motivated by the Lifshitz theory in solid state physics\cite{Horava},
allows the theory to be power-counting ultraviolet (UV) renormalizable with
the prescription to recover general relativity in the infrared (IR) limit.
Nevertheless, Ho\v{r}ava-Lifshitz (HL) theory is noncovariant. Another
proposal comes from Gravity's Rainbow which distorts the metric tensor around
and beyond the Planck scale\cite{MagSmo}. The basic ingredient is the
definition of two arbitrary functions: $g_{1}\left(  E/E_{\mathrm{Pl}}\right)
$ and $g_{2}\left(  E/E_{\mathrm{Pl}}\right)  $, which have the following
property%
\begin{equation}
\lim_{E/E_{\mathrm{Pl}}\rightarrow0}g_{1}\left(  E/E_{\mathrm{Pl}}\right)
=1\qquad\mathrm{and}\qquad\lim_{E/E_{\mathrm{Pl}}\rightarrow0}g_{2}\left(
E/E_{\mathrm{Pl}}\right)  =1.\label{lim}%
\end{equation}
$g_{1}\left(  E/E_{\mathrm{Pl}}\right)  $ and $g_{2}\left(  E/E_{\mathrm{Pl}%
}\right)  $ distort the metric in the following way. For a spherically
symmetric metric, we can define%
\begin{equation}
ds^{2}=-\frac{N^{2}\left(  r\right)  }{g_{1}^{2}\left(  E/E_{\mathrm{Pl}%
}\right)  }dt^{2}+\frac{dr^{2}}{\left(  1-\frac{b\left(  r\right)  }%
{r}\right)  g_{2}^{2}\left(  E/E_{\mathrm{Pl}}\right)  }+\frac{r^{2}}%
{g_{2}^{2}\left(  E/E_{\mathrm{Pl}}\right)  }\left(  d\theta^{2}+\sin
^{2}\theta d\phi^{2}\right)  ,\label{dS}%
\end{equation}
where $N\left(  r\right)  $ is known as the lapse function and $b\left(
r\right)  $ is the shape function. For black holes and wormholes, one has the
further condition $b\left(  r_{t}\right)  =r_{t}$ and therefore $r\in\left[
r_{t},+\infty\right)  $. This is not trivial, because for instance for Dark
Energy Stars, the radius $r\in\left[  0,+\infty\right)  $ for $b\left(
r\right)  $\cite{DBGL}. On the other hand for a FLRW metric, we can write%
\begin{equation}
ds^{2}=-\frac{N^{2}\left(  t\right)  }{g_{1}^{2}\left(  E/E_{\mathrm{Pl}%
}\right)  }dt^{2}+\frac{a^{2}\left(  t\right)  }{g_{2}^{2}\left(
E/E_{\mathrm{Pl}}\right)  }d\Omega_{3}^{2}~.\label{FRWMod}%
\end{equation}
Of course, ordinary gravity is recovered when $E/E_{\mathrm{Pl}}\rightarrow0
$. In a series of papers\cite{GaMa,GRw,RGBM}, it has been shown that Gravity's
Rainbow can keep under control UV divergences, at least to one loop. This
procedure has been widely tested on a spherically symmetric background. If one
considers perturbations of the metric (\ref{dS}), one finds that
Eq.(\ref{VEVO}) becomes%
\begin{equation}
\frac{g_{2}^{3}\left(  E/E_{\mathrm{Pl}}\right)  }{\tilde{V}}\frac
{\left\langle \Psi\left\vert \int_{\Sigma}d^{3}x\tilde{\Lambda}_{\Sigma
}\right\vert \Psi\right\rangle }{\left\langle \Psi|\Psi\right\rangle }%
=-\frac{\Lambda}{\kappa},\label{WDW1}%
\end{equation}
with%
\begin{equation}
\tilde{\Lambda}_{\Sigma}=\left(  2\kappa\right)  \frac{g_{1}^{2}\left(
E/E_{\mathrm{Pl}}\right)  }{g_{2}^{3}\left(  E/E_{\mathrm{Pl}}\right)  }%
\tilde{G}_{ijkl}\tilde{\pi}^{ij}\tilde{\pi}^{kl}\mathcal{-}\frac{\sqrt
{\tilde{g}}\tilde{R}}{\left(  2\kappa\right)  g_{2}\left(  E/E_{\mathrm{Pl}%
}\right)  }\!{}\!.\label{LambdaR}%
\end{equation}
The symbol \textquotedblleft$\sim$\textquotedblright\ indicates the quantity
computed in absence of rainbow's functions $g_{1}\left(  E/E_{\mathrm{Pl}%
}\right)  $ and $g_{2}\left(  E/E_{\mathrm{Pl}}\right)  $. Extracting the TT
tensor contribution from Eq.(\ref{WDW1}), we find that the total one loop
energy density for the graviton, or ZPE, induces a dynamical cosmological
constant whose form is%
\begin{equation}
\frac{\Lambda}{8\pi G}=-\frac{1}{3\pi^{2}}\sum_{i=1}^{2}\int_{E^{\ast}%
}^{+\infty}E_{i}g_{1}\left(  E/E_{\mathrm{Pl}}\right)  g_{2}\left(
E/E_{\mathrm{Pl}}\right)  \frac{d}{dE_{i}}\sqrt{\left(  \frac{E_{i}^{2}}%
{g_{2}^{2}\left(  E/E_{\mathrm{Pl}}\right)  }-m_{i}^{2}\left(  r\right)
\right)  ^{3}}dE_{i},\label{Lambda}%
\end{equation}
where we have defined two r-dependent effective masses $m_{1}^{2}\left(
r\right)  $ and $m_{2}^{2}\left(  r\right)  $, corresponding to the two
degrees of freedom of the graviton,%
\begin{equation}
\left\{
\begin{array}
[c]{c}%
m_{1}^{2}\left(  r\right)  =\frac{6}{r^{2}}\left(  1-\frac{b\left(  r\right)
}{r}\right)  +\frac{3}{2r^{2}}b^{\prime}\left(  r\right)  -\frac{3}{2r^{3}%
}b\left(  r\right) \\
\\
m_{2}^{2}\left(  r\right)  =\frac{6}{r^{2}}\left(  1-\frac{b\left(  r\right)
}{r}\right)  +\frac{1}{2r^{2}}b^{\prime}\left(  r\right)  +\frac{3}{2r^{3}%
}b\left(  r\right)
\end{array}
\right.  \quad\left(  r\equiv r\left(  x\right)  \right)  .\label{masses}%
\end{equation}
We have to observe that only appropriate choices of $g_{1}\left(
E/E_{\mathrm{Pl}}\right)  $ and $g_{2}\left(  E/E_{\mathrm{Pl}}\right)  $ lead
to a finite induced cosmological constant. For instance, if we choose%
\begin{equation}
g_{1}\left(  E/E_{\mathrm{Pl}}\right)  =1-\eta\left(  E/E_{\mathrm{Pl}%
}\right)  ^{n}\qquad\text{and}\qquad g_{2}\left(  E/E_{\mathrm{Pl}}\right)
=1,
\end{equation}
where $\eta$ is a dimensionless parameter and $n$ is an integer\cite{g1g2},
the ZPE of Eq.(\ref{Lambda}) diverges and therefore will be discarded. On the
other hand, if we choose%
\begin{equation}
g_{1}\left(  E/E_{\mathrm{Pl}}\right)  =(1+\beta\frac{E}{E_{\mathrm{Pl}}}%
)\exp(-\alpha\frac{E^{2}}{E_{\mathrm{Pl}}^{2}})\qquad g_{2}\left(
E/E_{\mathrm{Pl}}\right)  =1,\label{a)}%
\end{equation}
with $\alpha>0$ and $\beta\in%
\mathbb{R}
$, the ZPE of Eq.(\ref{Lambda}) is finite for the pure \textquotedblleft%
\textit{Gaussian}\textquotedblright\ choice, ($\beta=0$) but it does not work
correctly because it does not lead to a positive induced cosmological
constant. On the other hand the \textquotedblleft\textit{Non-Gaussian}%
\textquotedblright\ choice works correctly especially for a de Sitter, Anti de
Sitter and Minkowski background. To fix ideas, for a de Sitter background we
have $b\left(  r\right)  =$ $\Lambda_{dS}r^{3}/3$ with $\Lambda_{dS}>0$. Thus
Eq.(\ref{Lambda}) leads to the following behavior of $\Lambda/\left(  8\pi
G\right)  $%
\begin{equation}
\frac{\Lambda}{8\pi G}\simeq\left\{
\begin{array}
[c]{cc}%
-{\frac{4{\alpha}^{5/2}+3\sqrt{\pi}\beta{\alpha}^{2}}{4\pi^{2}{\alpha}^{9/2}}%
}E_{P}^{4} & x\rightarrow0\\
& \\
E_{P}^{4}\exp(-\alpha x^{2}) & x\rightarrow\infty
\end{array}
\right.  .
\end{equation}
By imposing that%
\begin{equation}
\beta=-{\frac{4\,}{3}}\sqrt{\frac{\alpha}{\pi}},\label{sm}%
\end{equation}
$\Lambda/\left(  8\pi G\right)  $ vanishes for small $x$ and therefore the
result is regular for every value of $x$, where $x=\sqrt{m_{0}^{2}\left(
r\right)  /E_{P}^{2}}$ and where $m_{0}^{2}\left(  r\right)  $ play the
r\^{o}le of an effective mass with%
\begin{equation}
m_{0}^{2}\left(  r\right)  =\frac{6}{r^{2}}-\Lambda_{dS}.\label{potentials}%
\end{equation}
This \textquotedblleft\textit{dynamical induced cosmological constant}%
\textquotedblright\ is finite for every choice of $r$. It is interesting to
note that the whole setting successfully applies also to the case of a naked
singularity\cite{RGBM} of the form (\ref{dS}) with $b\left(  r\right)  =-2MG$,
$M>0$. One finds that if we define%
\begin{equation}
x=\sqrt{\frac{m_{1}^{2}(r)}{E_{P}^{2}}}=\frac{1}{rE_{P}}\sqrt{6+\frac{15MG}%
{r}}\qquad\mathrm{and\qquad}y=\sqrt{\frac{m_{2}^{2}(r)}{E_{P}^{2}}}=\frac
{1}{rE_{P}}\sqrt{6+\frac{9MG}{r}},\label{xy}%
\end{equation}
then%
\begin{equation}
\lim_{\substack{x\rightarrow0 \\y\rightarrow0}}\frac{\Lambda}{8\pi GE_{P}^{4}%
}\simeq-\frac{4+3\sqrt{\pi/\alpha}~\beta}{4\pi^{2}\alpha^{2}}=0,
\end{equation}
if we set%
\begin{equation}
\beta=-\frac{4}{3}\sqrt{\frac{\alpha}{\pi}}.\label{beta}%
\end{equation}
On the other hand, when $r\rightarrow0$ or $M\rightarrow\infty$, one gets%
\begin{equation}
\frac{\Lambda}{8\pi GE_{P}^{4}}\simeq\frac{2}{3\pi^{2}\alpha r^{3}E_{P}^{2}%
}\left[  15MG\exp\left(  -\frac{\alpha15MG}{r^{3}E_{P}^{2}}\right)
+9MG\exp\left(  -\frac{\alpha9MG}{r^{3}E_{P}^{2}}\right)  \right]  .
\end{equation}
The case in which $M\rightarrow\infty$ is unphysical because it represents a
singularity which fills the whole universe. On the other hand the case in
which $r\rightarrow0$ represents a naked singularity which is no more
singular, at least from the ZPE point of view. Note that the choice (\ref{a)})
has been borrowed by Noncommutative Geometry (NCG). In NCG, one introduces a
deformation of the classical Liouville counting number of nodes%
\begin{equation}
dn=\frac{d^{3}\vec{x}d^{3}\vec{k}}{\left(  2\pi\right)  ^{3}}%
\end{equation}
with\cite{RGPN}%
\begin{equation}
dn_{i}=\frac{d^{3}\vec{x}d^{3}\vec{k}}{\left(  2\pi\right)  ^{3}}\exp\left(
-\frac{\theta}{4}k_{i}^{2}\right)  ,\label{moddn}%
\end{equation}
where%
\begin{equation}
k_{i}^{2}=E_{i}^{2}-m_{i}^{2}\left(  r\right)  \quad i=1,2.
\end{equation}
$\theta$ is the parameter encoding the noncommutativity of spacetime
represented by the commutator $\left[  \mathbf{x}^{\mu},\mathbf{x}^{\nu
}\right]  =i\,\theta^{\mu\nu}$. $\theta^{\mu\nu}$ is an antisymmetric matrix
which determines the fundamental discretization of spacetime. However, since
Gravity's Rainbow depends on $g_{1}\left(  E/E_{\mathrm{Pl}}\right)  $ and
$g_{2}\left(  E/E_{\mathrm{Pl}}\right)  $ and not only by one single parameter
like $\theta$, it appears to be more flexible in describing ZPE calculations.
This flexibility appears more evident when one tries to make a bridge between
Gravity's Rainbow and HL theory. Indeed it is possible to create a
correspondence between them at least for a FLRW metric and for a spherically
symmetric background\cite{RGENS}. This is possible when%
\begin{align}
&  g_{1}\left(  E/E_{P}\right)  \equiv g_{1}\left(  E\left(  a\left(
t\right)  \right)  /E_{P}\right) \nonumber\\
&  g_{2}\left(  E/E_{P}\right)  \equiv g_{2}\left(  E\left(  a\left(
t\right)  \right)  /E_{P}\right)
\end{align}
for a FLRW metric and%
\begin{align}
&  g_{1}\left(  E/E_{P}\right)  \equiv g_{1}\left(  E\left(  b\left(
r\right)  \right)  /E_{P}\right) \nonumber\\
&  g_{2}\left(  E/E_{P}\right)  \equiv g_{2}\left(  E\left(  b\left(
r\right)  \right)  /E_{P}\right)  ,
\end{align}
for a spherically symmetric metric. This interesting connection gives a hint
on why Gravity's Rainbow produces finite results, at least in one loop calculations.


\begin{thebibliography}{99}                                                                                               %
\bibitem {DeWitt}B. S. DeWitt, \textsl{Phys. Rev.} \textbf{160}, 1113 (1967).

\bibitem {RemoHL}R. Garattini, \textsl{Phys. Rev. }\textbf{D 86}, 123507
(2012); ArXiv:0912.0136 [gr-qc].

\bibitem {RGMdL}R. Garattini and M. De Laurentis, in preparation.

\bibitem {Remo}R.~Garattini, TSPU Vestnik \textbf{44 N 7}, 72 (2004),
ArXiv:gr-qc/0409016; R.~Garattini, \textsl{J. Phys.} \textbf{A 39}, 6393
(2006), ArXiv:gr-qc/0510061; R.~Garattini, \textsl{J. Phys. Conf. Ser.}
\textbf{33}, 215 (2006), ArXiv:gr-qc/0510062. R. Garattini, \textsl{J. Phys.}
\textbf{A 41} (2008) 164057, ArXiv:0712.3246 [gr-qc]; R. Garattini,
\textsl{AIP Conf. Proc.} \textbf{1241} 866 (2010), ArXiv: 0911.2393 [gr-qc].
S. Capozziello and R. Garattini, \textsl{Class.Quant.Grav.} \textbf{24}, 1627
(2007); ArXiv: 0702075 [gr-qc].

\bibitem {Horava}P. H\v{o}rava, \textsl{JHEP}, 0903, \textbf{020} (2009).
ArXiv: 0812.4287 [hep-th]; P. H\v{o}rava, \textsl{Phys. Rev. Lett.}
\textbf{102}, 161301 (2009) ArXiv: 0902.3657 [hep-th]. P. H\v{o}rava,
\textsl{Phys. Rev.} \textbf{D 79}, 084008 (2009). ArXiv: 0901.3775 [hep-th].

\bibitem {MagSmo}J.~Magueijo and L.~Smolin, \textsl{Class.\ Quant.\ Grav.}%
\ \textbf{21}, 1725 (2004). ArXiv: 0305055 [gr-qc].

\bibitem {DBGL}A.DeBenedictis, R. Garattini and F. S.N. Lobo,
\textsl{Phys.Rev. }\textbf{D 78}, 104003 (2008), ArXiv:0808.0839 [gr-qc].

\bibitem {GaMa}R.~Garattini and G.~Mandanici, \textsl{Phys. Rev.} \textbf{D
83}, 084021 (2011); ArXiv:1102.3803 [gr-qc].

\bibitem {GRw}R.~Garattini, \textsl{Phys.\ Lett.}\ \textbf{B} \textbf{685},
329 (2010), ArXiv: 0902.3927 [gr-qc]. R.~Garattini and G.~Mandanici,
\textsl{Phys. Rev.} \textbf{D 85}, 023507 (2012); ArXiv:1109.6563 [gr-qc].
R.~Garattini, JCAP \textbf{1306}, 017 (2013), ArXiv: 1210.7760 [gr-qc].
R.~Garattini and F.~S.~N.~Lobo, \textsl{Phys.\ Rev.}\ \textbf{D} \textbf{85},
024043 (2012), ArXiv:1111.5729 [gr-qc]. R.~Garattini,
\textsl{Int.\ J.\ Mod.\ Phys.\ Conf.\ Ser.}\ \textbf{14}, 326 (2012),
ArXiv:1112.1630 [gr-qc]. R. Garattini and F.S.N. Lobo, \textsl{Eur. Phys. J.}
\textbf{C 74} (2014), ArXiv:1303.5566 [gr-qc]. R. Garattini and B.
Majumder,\textsl{\ Nucl. Phys.} \textbf{B} \textbf{883} (2014),
ArXiv:1305.3390 [gr-qc]. R. Garattini and M. Sakellariadou, \textsl{Phys.
Rev.} \textbf{D} \textbf{90} 043521 (2014), ArXiv:1212.4987 [gr-qc].

\bibitem {RGBM}R. Garattini and B. Majumder, \textsl{Nucl. Phys.} \textbf{B}
\textbf{884, }(2014), ArXiv:1311.1747 [gr-qc].

\bibitem {g1g2}Y. Ling, \textsl{JCAP} \textbf{17}, 708 (2007), gr-qc/0609129;
Y. Ling, X. Li and H. Zhang, \textsl{Mod.Phys.Lett. }\textbf{A 22}, 2749
(2007), gr-qc/0512084; Y. Ling, B. Hu and X. Li, \textsl{Phys. Rev.} \textbf{D
73}, 087702 (2006), gr-qc/0512083.

\bibitem {RGPN}R. Garattini and P. Nicolini, \textsl{Phys. Rev.} \textbf{D
83}, 064021 (2011); ArXiv:1006.5418 [gr-qc].

\bibitem {RGENS}R. Garattini and E. N. Saridakis, \textit{Gravity's Rainbow: a
bridge towards Horava-Lifshitz gravity;} ArXiv:1411.7257.
\end{thebibliography}
\end{document}